# Spreadsheet modelling for solving combinatorial problems: The vendor selection problem

Pandelis G. Ipsilandis
Project Management Department, Technological Education Institute of Larissa
Larissa Greece, 41110, E-mail: ipsil@teilar.gr

**ABSTRACT**

Spreadsheets have grown up and became very powerful and easy to use tools in applying analytical techniques for solving business problems. Operations managers, production managers, planners and schedulers can work with them in developing solid and practical Do-It-Yourself Decision Support Systems. Small and Medium size organizations, can apply OR methodologies without the presence of specialized software and trained personnel, which in many cases cannot afford anyway. This paper examines an efficient approach in solving combinatorial programming problems with the use of spreadsheets. A practical application, which demonstrates the approach, concerns the development of a spreadsheet-based DSS for the Multi Item Procurement Problem with Fixed Vendor Cost. The DSS has been build using exclusively standard spreadsheet feature and can solve real problems of substantial size. The benefits and limitations of the approach are also discussed.

## 1 INTRODUCTION

Since their introduction in the early eighties electronic spreadsheet programs such as Excel, Lotus 1-2-3 and Quattro Pro have grown to be the most common tool managers use to model and analyze quantitative problems. The latest versions of spreadsheets contain powerful analytical tools accessible through a user-friendly interface that provide end users with such computing power, we could only dream several years ago. In today's business world an unprecedented number of managers are familiar with quantitative modelling tools available at their fingertips through the spreadsheet software in their desktop computers. Furthermore, the great majority of them understand numbers, systems and relationships [Ashley, (1995)]. Therefore although they may not be very well versed in the techniques of operations research, they possess the required fundamental skills for developing end-user decision support tools that make use of OR/MS principles and methodologies.

Historically, optimization problems were typically solved using special purpose optimization software packages such as LINDO, MATLAB, SAS etc. [Novak, et al., (2003)] which is widely available in the academic world, but is not commonly found in businesses since it is not considered part of the everyday business tools that managers use. On the other hand, the spreadsheet analysis tools including the optimizer known as "solver" which is currently the most readily available general-purpose optimization modelling system are available to approximately 35 million users of office productivity software worldwide. Their widespread availability has spawned many optimization applications in both the private and public sectors [Fylstra et al (1998)]. Many users without a mathematical background became capable of developing useful and substantive decision support tools utilizing OR methodologies [Roy et al (1989)].

95

# Spreadsheet modelling for solving combinatorial problems: vendor selection
Pandelis G. Ipsilandis

The relevant research literature addressing spreadsheet optimization and end-user spreadsheet modelling is quite extensive in the last years. Both academics and business practitioners report on successful applications of end user decision modelling with the help of spreadsheets.

Several authors [Barlow (1999), Thiriez (2001), Hesse (1997)] propose modelling schemes for almost all the familiar OR problems and methodologies such as linear programming, transportation problems, decision theory, dynamic programming, simulation etc. and their applications, which utilize modern spreadsheet features such as mathematical and reference functions, random number generators, and the "solver" facility. Although, they present spreadsheet templates for OR problem as an alternative to OR/MS software, yet the proposed models are intended mostly for educational use in a classroom environment, since they refer to very well structured problems, of limited size in many cases; therefore not too suitable for real life applications.

More interesting spreadsheet models are the ones that refer to less structured problems such as solving discrete state and discrete stage dynamic programs [Parlar (1989)] and stochastic programming with multiple decision criteria applications [Novak (2003)] as well as spreadsheet models that have been used to solve real life problems in a wide diversity of application areas such as: using multivariate constrained optimization in production scheduling [Osman (1992)], university library staff scheduling using integer programming techniques [Ashley (1995)], optimization of bank customer redistribution through a multi stage decision making model [Chu (1995)],.queuing decomposition models in production work flow [Enns (1999)], marketing decision making models [Albers (2000)], integrated business planning including sales forecasting, production planning and cash budgeting [Chien (2000)], design of the workload of paced or linked assembly lines [Johnson (2002)], facility location problems [Pearson (2003)], to mention some of the most important during the last years.

This paper presents an end-user spreadsheet-based model for solving a combinatorial integer-programming problem referring to the optimum selection of vendors in a multi-item procurement problem. Since the excel solver works very poorly for problems with many binary variables, the spreadsheet model is based on other spreadsheet features such as extensive use of mathematical and reference functions, and data tables for "smart" enumeration of alternative solutions. In this sense it does not follow a formal OR model structure; instead, the model uses typical business-like layout of the problems parameters, in a way that facilitates the managerial presentation and the analysis of the results.

The remainder of this paper is organized as follows: Section 2 describes the real case of a multi item procurement problem exploring the problem's parameters, the difficulty in estimating certain cost parameters and some of the managerial concerns involved in the decision making process. Section 3 provides a formal definition of the multi-item procurement problem, and its formulation as an integer programming optimization problem. It also provides some insight to the nature of the solutions of the problem and references to heuristic algorithms and other approximate methods, which can be used to produce near-optimum solutions. Section 4 describes the spreadsheet model used to solve the problem. Finally, Section 6 contains a summary of the conclusions derived.

## 2 THE MULTI ITEM VENDOR SELECTION (MIVS) PROBLEM

Vendor selection and procurement policies are very important since they can affect dramatically the cost, the quality and the promised delivery time of deliverables of a specific project or operation. The quantity and variety of vendor selection models published in the operations research, production management and other related literature [Degraeve et al. (2000)] reflect not only the complexity but also the financial importance of these decisions. A special type of procurement problem is the Multi Item Vendor

Spreadsheet modelling for solving combinatorial problems: vendor selection
Pandelis G. Ipsilandis

Selection (MIVS) problem which refers to procurement decisions involving the acquisition of a specific set of discrete items, all of which can be supplied by many different vendors, with no interdependencies (i.e. functional, price, demand etc.) or other variations (e.g. quality) between them, where the procurement decision is solely cost-based. A typical application of this problem is an open public call of tenders for the procurement of a list of items where interested vendors submit their bids. The problem is not trivial because the cost to the procurer includes a certain fixed handling cost for each selected vendor besides the direct acquisition cost. This cost comprises components such as: legal review, contract administration, communications, invoicing, receipt of materials, bill of delivery processing, etc. The procurer bears this fixed cost regardless of the quantity of items the specific supplier delivers. It is obvious that if the vendor selection process results in choosing many vendors for the supply of the given set of items, the total supplier handling cost increases, leading consequently to increases in the total cost. Hence, procurement managers have a tendency to identify and select a limited number of suppliers, in order to keep the procurement administrative cost low. However as markets become more open and firms move from single sourcing to competitive tendering [Reid (1989)], while at the same time organizations such as the World Trade Organization (WTO) and the European Union (EU) issue directives to ensure that public procurement policies are more open, transparent, doubtless but at the same time efficient [Hoekman (1998); Ferguson et al. (1995)], the issue of selecting the optimum number of vendors in a multi-item procurement becomes very important in order to maintain cost-efficiency. The following is a real life case of the MIVS problem.

**2.1 A Case of The MIVS Problem: Procurement of Library Books**

The expansion of regional school libraries was set as a top priority goal of the Greek ministry of education in the late 90s. In the period 1998-2000, a number of such projects were initiated at different regions in the country, the management of which was assigned to local universities because of their expertise in library operations. Each project involved a number of activities such as: preparing the library space, organizing the library functions, hiring and training staff, installing IT services etc. The Technological Education Institute of Larissa managed one of these projects in the region of Central Greece, which involved, among other tasks, the acquisition of approximately 4.000 titles of books corresponding to a budget over 200.000 euros.
The project manager was obliged to follow all financial regulations and policies concerning public procurement imposed by the EU and the national government both of which funded the project. Ten bids by major book suppliers were submitted, following an open call for tenders. Due to the nature of the products to be purchased, no quality or other differences existed among the different vendors, therefore, the only selection criterion that could be used is the price offered by each vendor. Initially, the project manager considered two different supplier selection alternatives:

*Alternative I*   *Select a single vendor based on the least total purchasing cost.*

*Alternative II*   *Buy each book from the vendor who offered the lowest price for it.*

The project manager followed alternative II because it seemed reasonable, transparent, non-questionable, and easy to arrive at a justified decision. In addition, prevents a result where one firm gets the entire order based on a marginal cost difference, while other vendors could have offered lower prices for books in specific domains, which would lead to lower procurement cost. Figure 1 summarizes the results of the vendor selection policy followed. Nine vendors were selected with a minimum purchasing cost of 220.736 euros.

| Number of vendors submitted bids: 10 | | Number of vendors selected: 9 |
|---|---|---|
| Vendor | Number of Books supplied | Purchasing cost (€) |

Spreadsheet modelling for solving combinatorial problems: vendor selection
Pandelis G. Ipsilandis

| # 1 | 260 | 10.118 |
|---|---|---|
| # 2 | 140 | 4.008 |
| # 3 | 50 | 2.647 |
| # 4 | 160 | 8.075 |
| # 5 | 940 | 48.072 |
| # 7 | 50 | 2.469 |
| # 8 | 20 | 1.094 |
| # 9 | 1.610 | 66.847 |
| # 10 | 840 | 77.406 |
| Total | 4.070 titles | 220.736 € |

**Figure 1** – Results of procurement policy "Buy each book at the lowest price"

Although, alternative II results in the minimum total procurement cost, the total cost is not minimized. As it was discovered during the course of the project, a relatively high vendor handling cost existed, which was initially hidden. Although it is difficult to arrive at an objective estimation of this cost, a rough calculation, put it in the range of 3.600 € per vendor (figure 2), This vendor handling cost is paid for every vendor selected, even in the case where only a small number of items are assigned to the specific vendor, making this problem a typical MIVS problem.

| Personnel Cost based on person-days | Proj. Mnger | Libra-rian | Acctg Admin. | Legal Conslt. | Proj. Admin. |
|---|---|---|---|---|---|
| Proposal examination – Data verification and correction | | 2 | | 0,2 | 0,5 |
| Contract preparation – Financial/Legal Review | 0,3 | | 1 | 0,5 | 1 |
| Receiving – | | 2,5 | | | 1 |
| Payment preparation | 0,3 | | 2 | | 2 |
| Total person days per vendor | 0,6 | 4,5 | 3 | 0,7 | 4,5 |
| Total personnel cost: 13,3 days x 200 € per day       = 2.660 € | | | | | |
| Other cost (mailing etc)                              = 1.000 € | | | | | |
| Fixed Vendor Handling Cost      (approximately)       = 3.600 € | | | | | |

**Figure 2** – Analysis of the fixed vendor handling cost

Taking into consideration the fixed vendor handling cost, the total cost to the project is as follows:

Total book acquisition cost = Procurement cost + Fixed vendor handling cost

= 220.736 + 9 x 3.600 = 253.136 €

Obviously the policy that was followed is not optimal. Selecting fewer suppliers could probably result in a lower cost. As it will be explained in the next section, where a formal definition of the MIVS problem is given, obtaining an optimal solution to the problem is not a trivial task. A spreadsheet model can be used to assist managers in identifying the optimal solution to the problem.

## 3 AN INTEGER PROGRAMMING FORMULATION OF THE MIVS PROBLEM

### 3.1 Formal Problem Definition

*3.1.1   Assumptions*

The multi-item vendor selection (MIVS) problem under consideration is defined as the problem of minimizing the total cost of procuring *m* different items from *n* given vendors. The following operational assumptions are made:

i   Each item is bought from a single vendor (no splitting of the total demanded quantity of a given item among different vendors is allowed). Therefore without loss of generality we can assume that only one unit (i.e. the total demanded quantity) of each item is to be obtained and vendors bid for it if they have the capacity to supply it.
ii  No significant differences exist between vendors in terms of qualitative factors such as: product quality, vendor reliability, delivery time etc.
iii There are no price, demand or functional dependencies of any kind between any of the m items. Without loss of generality we can assume that all *n* vendors offer all *m* items at a cost of $p_{ij}$ *(i=1,...,m, j=1,...,n)*, since the case of vendor j not offering item i can be handled by setting the corresponding cost $p_{ij}$ very high.
iv  The selection of any vendor j, imposes a fixed cost $c_j$ to the procurer. This cost is incurred regardless of the number of items that are bought from the specific vendor.

These assumptions are compatible with many situations in public procurement dealings. In these cases a minimum set of requirements is set regarding product features and quality, vendor reliability, expected delivery times etc., a public call of tenders is announced and the buying decision are mainly cost based.

**Mathematical Model**

Based on the formal definition given above, the MIVS problem can be formulated as an integer linear programming model as follows:
Let *i= 1,,2,...,m* denote the items and *j=1,2,...,n* the vendors. Define $x_{ij}$ and $y_j$ to be indicator variables where

$$x_{ij} = \begin{cases} 1, & \text{if item i is bought from vendor j} \\ 0, & \text{otherwise} \end{cases}$$

and

$$y_j = Minimum\{\sum_{i=1}^{m} x_{ij}, 1\}, \quad i = 1,...,m, \quad j = 1,...,n.$$

In other words, $y_j = 1$, if at least one item is bought from the $j^{th}$ vendor, otherwise $y_j = 0$.
Let $p_{ij}$ be the price of the $i^{th}$ item charged by the $j^{th}$ vendor and $c_j$ the fixed cost to the procurer if at least one item is bought from the $j^{th}$ vendor.
Then the problem of optimum procurement can be written in linear form as follows:

$$\text{Minimize} \quad C = \sum_{i=1}^{m}\sum_{j=1}^{n} p_{ij} * x_{ij} + \sum_{j=1}^{n} c_j * y_j \tag{1}$$

subject to

$$\sum_{j=1}^{n} x_{ij} = 1, \quad \forall \ i = 1,..,m \tag{2}$$

$$x_{ij} \leq y_j \quad \forall \ i = 1,...,m, \quad j = 1,...,n \tag{3}$$

$$x_{ij}, y_j = 0,1 \quad \forall \ i=1,...,m, \quad j=1,..,n \tag{4}$$

Note that the set of constraints (3) and (4) replaces the definition of the indicator variable $y_j$ given earlier. If for a given vendor *j*, at least one item is selected then one of the $x_{ij}$'s will be equal to one, thus forcing $y_j$ to get a value of at least 1. In addition because of the minimization of the objective function the $y_j$'s can never assume values greater than 0 if no item is supplied by vendor *j*.

**3.2 Exploring the Solution Space of the MIVS Problem**

Spreadsheet modelling for solving combinatorial problems: vendor selection
Pandelis G. Ipsilandis

The mathematical model for the MIVS problem is similar to the facility location problem, a well-known NP-complete combinatorial problem [Akinc (1993)], and therefore it is rather unlikely to find an efficient algorithm for solving it [Cornuejols, et. al. (1997)]. The solution space of the MIVS problem is combinatorial in nature; therefore the number of feasible variable combinations to be searched by an (explicit or implicit) enumeration algorithm (such as branch and bound) in order to locate the optimum solution is enormous. Based on the mathematical model defined in 3.1 and especially considering constraints (2), it is easily concluded that the maximum number of feasible variable combinations (and so candidate solutions) is $n^m$. Even for moderate values of $n$ (number of vendors) and $m$ (number of items to be purchased) the solution space may grow into very large numbers.

Heuristic algorithms for this sort of problems have been developed based on the observation that for any $k$ given vendors it is easy to identify the minimum cost solution by assigning each item to the vendor (out of the set of the $k$ given vendors) who offers the lowest price for this specific item. The spreadsheet model, which is described in the next section, is based on this observation.

Consider a vector $Y^r = (y^r_1, y^r_2, \ldots, y^r_j, \ldots, y^r_n)$, the elements of which take binary values. We can call $Y^r$ a vendor selection vector, since each vector $Y^r$ corresponds to a specific selection of k vendors (those with corresponding $y^r_j = 1$).

Then for a given $Y_r$, an associated minimum cost solutions is defined as follows:

$S_r = \{ x_{ij} : x_{ij}=1$ if $c_{ij} = \min_i(x_{ij})$ and $y^r_j=1$; $x_{ij}=0$ otherwise $\forall I=1,\ldots,m$ and $\forall j=1,\ldots,n\}$

The number of the solutions $S_r$ for any MIVS problem is equal to $2^n-1$ and is independent of the number of items $m$. This can be shown easily by considering all possible combinations of n variables each taking only the value of 0 or 1 and excluding the one combination where all variables take the value of 0 (i.e. at least one vendor must be selected). Each solution $S_r$ is a candidate for optimizing the total cost. The optimum solution for the problem can be found by comparing the cost of all the $S_r$ solutions.

Figure 3 illustrates the above for a problem with 5 vendors and 9 products. The table on the left includes the prices offered by each vendor plus the vendor fixed cost. The table on the right presents a solution where vendors 2, 3 and 5 are selected. The total cost is the acquisition cost (minimum price for each item among the specific vendors) plus the fixed cost of all selected vendors.

**Price Matrix (Pij)**
*Suppliers*

|  | S1 | S2 | S3 | S4 | S5 |
|---|---|---|---|---|---|
| P1 | 19 | 13 | 11 | 12 | 12 |
| P2 | 19 | 17 | 16 | 13 | 10 |
| P3 | 15 | 14 | 21 | 18 | 11 |
| P4 | 16 | 23 | 24 | 23 | 14 |
| P5 | 23 | 11 | 16 | 11 | 24 |
| P6 | 18 | 16 | 20 | 18 | 11 |
| P7 | 22 | 18 | 22 | 20 | 11 |
| P8 | 23 | 24 | 16 | 14 | 22 |
| P9 | 12 | 10 | 10 | 14 | 16 |

Fixed Cost Cj: 10 | 13 | 15 | 8 | 11

**Vendors Selected**

Yj= 0 | 1 | 1 | 0 | 1

|  | S1 | S2 | S3 | S4 | S5 | Min Pi |
|---|---|---|---|---|---|---|
| P1 |  | 13 | 11 |  | 12 | 11 |
| P2 |  | 17 | 16 |  | 10 | 10 |
| P3 |  | 14 | 21 |  | 11 | 11 |
| P4 |  | 23 | 24 |  | 14 | 14 |
| P5 |  | 11 | 16 |  | 24 | 11 |
| P6 |  | 16 | 20 |  | 11 | 11 |
| P7 |  | 18 | 22 |  | 11 | 11 |
| P8 |  | 24 | 16 |  | 22 | 16 |
| P9 |  | 10 | 10 |  | 16 | 10 |

Acuisition Cost: 105

Cj: | 13 | 15 | | 11 | 39 Vendor Fixed Cost

Total cost of solution: 144
Number of vendors in solution: 3

**Figure 3** – An example of a MIVS problem

**4 A SPREADSHEET MODEL FOR THE MIVS PROBLEM**

Spreadsheet modelling for solving combinatorial problems: vendor selection
Pandelis G. Ipsilandis

One of the advantages of spreadsheet modelling is that the problem data and parameters can be organized in a business-like format, avoiding the use of mathematical notation and special input formatting. The spreadsheet model used for solving the MIVS problem is explained analytically through a smaller size demonstration example shown in figure 3 with five vendors and nine items. For any given vendor selection shown by setting corresponding cells equal to 1 (i.e. vendors 2, 3, and 5 are selected in figure 3) the resulting solution is trivial (shaded cells correspond to minimum price items).

**4.1 Organization of the Problem's Data**

The parameters for the MIVS problem consist of an $m \times n$ matrix of the prices $p_{ij}$ (i.e. price of item $i$ offered by vendor $j$), and the $1 \times n$ vector of the fixed costs for each of the $n$ vectors. A spreadsheet called "problem data" holds the values of the problem's parameters as shown in figure 4.

|   | A | B | C | D | E | F | G | H | I |
|---|---|---|---|---|---|---|---|---|---|
| 1 |   |   |   |   | MIVS Problem - Values of Problem Parameters |   |   |   |   |
| 2 |   |   |   |   | Price Matrix (Pij) |   |   |   |   |
| 3 |   |   |   |   |   | Suppliers |   |   |   |
| 4 |   |   |   |   | S1 | S2 | S3 | S4 | S5 |
| 5 |   |   |   | P1 | 19 | 13 | 11 | 12 | 12 |
| 6 |   |   |   | P2 | 19 | 17 | 16 | 13 | 10 |
| 7 |   |   |   | P3 | 15 | 14 | 21 | 18 | 11 |
| 8 |   |   Products | P4 | 16 | 23 | 24 | 23 | 14 |
| 9 |   |   |   | P5 | 23 | 11 | 16 | 11 | 24 |
| 10 |   |   |   | P6 | 18 | 16 | 20 | 18 | 11 |
| 11 |   |   |   | P7 | 22 | 18 | 22 | 20 | 11 |
| 12 |   |   |   | P8 | 23 | 24 | 16 | 14 | 22 |
| 13 |   |   |   | P9 | 12 | 10 | 10 | 14 | 16 |
| 14 |   |   |   |   |   |   |   |   |   |
| 15 |   | Vendor Fixed Cost | Cj | 10 | 13 | 15 | 8 | 11 |

Problem Data / Opt Sol Enur

**Figure 4** – The MIVS problem's parameters

**4.2 Determining a Problem Solution**

As explained in 3.2, for any given vendor selection vector $Y^r = (y^r_1, y^r_2, \ldots, y^r_j, \ldots, y^r_n)$, corresponding to a choice of k specific vendors, the solution to the problem is trivial. It is found by obtaining each item from the vendors offering the lowest price for it. For a problem involving $n$ vendors the number of all possible $Y^r$ vectors is $2^n-1$. Since each of these vectors corresponds to a binary sequence of n binary digits it is easy to map the sequence $\{1, 2, 3, \ldots, 2^n-1\}$ to vectors matching the binary representation of each number in the list as follows:

$$\begin{array}{lll}
\text{Vector} & (\text{element 1, 2, 3,} & \text{n-2, n-1, n}) \\
Y^1 & \rightarrow & \{0, 0, 0, \quad , 0, 0, 1\} \\
Y^2 & \rightarrow & \{0, 0, 0, \quad , 0, 1, 0\} \\
Y^3 & \rightarrow & \{0, 0, 0, \quad , 0, 1, 1\} \quad \text{etc.}
\end{array}$$

Therefore one can produce a list of $2^n-1$ solutions, each corresponding to an integer $r$ between 1 and $2^n-1$. Figure 5 shows the spreadsheet model, which accepts as input a number $r$ and produces the $r^{th}$ solution corresponding to a given combination of vendors determined by the binary representation of $r$.

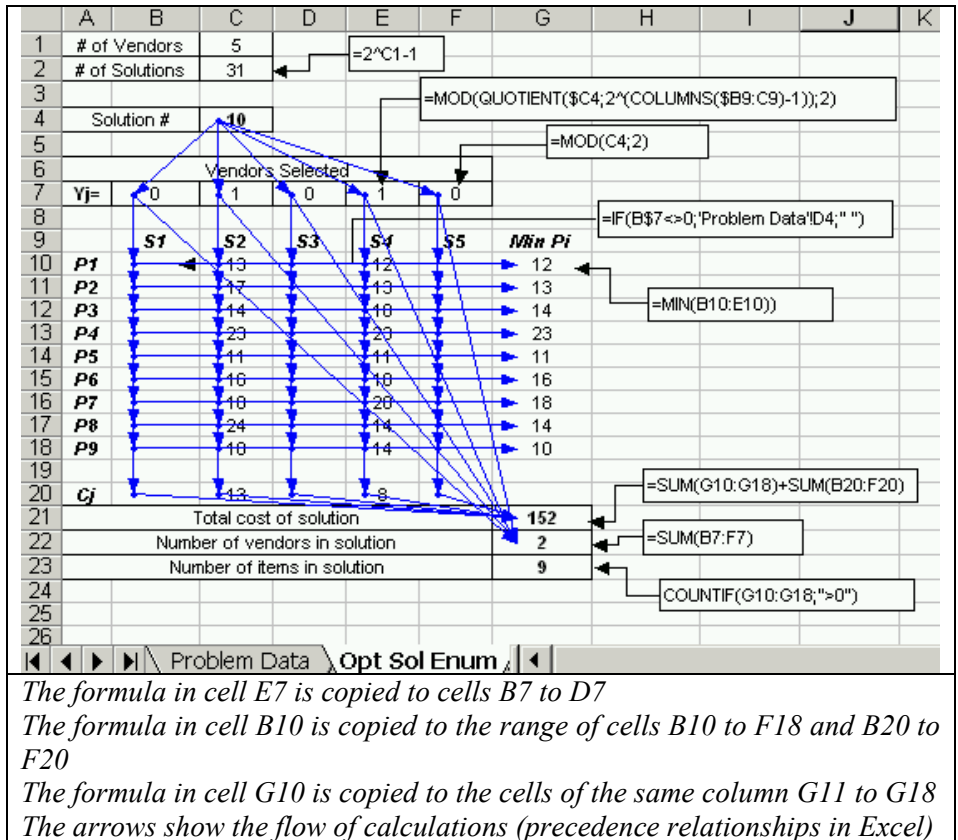

*The formula in cell E7 is copied to cells B7 to D7*
*The formula in cell B10 is copied to the range of cells B10 to F18 and B20 to F20*
*The formula in cell G10 is copied to the cells of the same column G11 to G18*
*The arrows show the flow of calculations (precedence relationships in Excel)*

**Figure 5** – Producing the $r^{th}$ solution to the MIVS problem

The solution sequence number $r$ is entered in cell *C4*.
The formulas in cells *B7* through *F7* convert this number in a binary representation of 5 digits. For the specific example the binary notation of number 10, in a 5-bit form is 01010, corresponding to selection of vendors 2 and 4.
The cell formulas in the range *B10* to *F18* bring from the original price matrix in the spreadsheet *problem data* (see figure 4) only the values for those columns that correspond to the specific choice of vendors associated with the $r^{th}$ solution.
Similar formulas bring the associated vendor fixed cost in cells *B20* to *F20*.
The minimum price for each item is computed in column G and the results for the $r^{th}$ solution is summarized in cells *G21*, *G22* and *G23*.
The computation of the number of items in the $r^{th}$ solution is necessary in the general case where some vendors do not offer all the *m* items.

**4.3 Finding the Optimum Solution**

If one tries all different values of the parameter *r* in the model described in the previous section, the optimum solution can be found by comparison of the corresponding cost of each solution. However the number of the solutions to be checked is very large even for moderate values of *n*. The Excel data-table feature automatically tabulates the results of complicated calculations, which cannot be expressed analytically, for different values of one or two input parameters. For the MIVS problem one can consider the solution sequence number *r* as the data-table input parameters and ask Excel to tabulate the values of the results shown in figure 5 (Cost, number of vendors, and number of items) for all possible values of *r*. Figure 6 demonstrates the organization of the data-table for the tabulation of the results.

Spreadsheet modelling for solving combinatorial problems: vendor selection
Pandelis G. Ipsilandis

**Figure 6** – MIVS problem: Finding the optimum solution

For the problem with five vendors there are 31 ($2^5$-1) possible choices for selecting 1, 2, 3, 4 or 5 vendors. The list of values 1 to 31 is entered in cells *M8* to *M38*. Formulas pointing to the cells holding the calculations of the results to be tabulated are entered in the top raw cells N7 to P7. Using the Table command in the Data drop down menu, and setting cell C4 as the column input cell, Excel automatically calculates and tabulates the results of the formulas entered in cells N7 to P7 when cell C4 takes every value in column M.

Following the tabulation of the results, the formulas in cells M3 to P3 calculate the optimum cost and determine the optimum solution to the problem.

**4.4 Results for the MIVS Problem: Procurement of Library Books**

The model described in the previous section was used in the case described in section2.1 for the optimization of the number of vendors for the procurement of library books. Figure 7 shows the organization of the problem's data and figure 8 the spreadsheet optimization.

**Figure 7** – Price and Fixed Cost Data for the MIVS problem

The dimensions of the price matrix were 4070 rows (each row representing a different book title) by 10 columns (each column corresponds to a vendor). For certain books in the list multiple copies were asked in the tender. Without any loss of generality, the definition of the item in the library book procurement MIVS problem each item is defined as the total quantity of each different book title, and the associated price of the item represented the cost of buying the asked quantity at the offered price. For example, item 1 in figure 7 corresponds to 100 copies of the specific book title.

The optimization results (figure 8) show that minimum cost is achieved if 3 vendors are selected (vendors 5, 9 and 10). The total cost of the optimum solution is 240.816 euro's, which is broken down to 230.016 € cost of the books and 10.800€ fixed vendor cost. If the suggested procurement policy were followed the corresponding cost would exceed 250.000€.

The spreadsheet model informs the user about which books are to be bought from each vendor and at what price, and allows "what-if" questions to be asked in order to further analyze the results. For example, the user can easily find suboptimum solutions corresponding to a given number of vendors in order to see what is the additional cost involved by increasing or decreasing the number of vendors. Also in cases where some books are offered by very few vendors, or not offered at all, the model allows the user to select a suboptimum solution by putting restrictions to the number of items to be purchased.

Figure 8 – Spreadsheet optimization of the MIVS problem

**4.5 Spreadsheet Modelling Limitations for Combinatorial Problems**

*Size limitations*

The spreadsheet model described in the previous sections performs very well as long as the size of the problem is such that the tabulation of the results produced by direct enumeration fit within the size of an excel spreadsheet. Given that the data table which holds the results must have $2^n-1$ rows to accommodate all possible vendor selection combinations, the latest version of an excel sheet with 65536 rows can accommodate problems with a maximum of 16 vendors. Although this may not present a practical problem as in many cases the number of vendors is well below this upper limit, it is still a limitation. A way to overcome it is by decomposing the problem into smaller sub-problems. However in combinatorial problems the number of solutions is growing explosively; for each additional vendor considered the number of solutions to be examined doubles.

*Computing time*

Combinatorial problems require excessive computing power and CPU time even with today's computers. Table 1 shows the CPU time for solving MIVS problems of different sizes on a personal computer with Intel Pentium 800Mhz processor.

| Computation time as a function of the MIVS problem size | | | |
|---|---|---|---|
| | Number of items: *m* | | |
| Number of Vendors: *n* | 100 | 200 | 400 |
| 10 | 5" | 10" | 22" |
| 12 | 27" | 45" | 1', 45" |
| 14 | 2', 05" | 3', 30" | 7', 45" |

**Table 1** – CPU time for obtaining an optimal solution to the MIVS problem

The required computation time is considerable for problems of larger size. Nonetheless np-complete combinatorial problems are problems, which are considered hard to solve, and no one could dream producing an optimum solution to this kind of problems using a personal computer several years ago. Empirical evidence drawn by the results in table 1 suggest that the computation time is almost linear in terms of the number of the parameter *m* (number of items in the problem), while it shows an exponential growth in terms of the parameter *n* (number of vendors).

*Why not using the Excel Solver*

Excel's built-in Solver is very powerful optimization tool but it performs very well only in linear problems. It presents problems in cases of nonlinear problems and it fails most of the time in problems with non-continuous variables. Even though the solver allows the modeler to specify that some variables are binary, or to "solve" nonlinear problems the results are not necessary reliable [Thiriez (2001)]. For the MIVS problem the solver failed to produce an optimum solution even for small demonstration problems like the one described in figure 3.

**5 CONCLUSIONS**

Spreadsheets have progressed well beyond the number crunching and graphics presentation into the problem solving and decision tool arena. In this paper we analyzed the use of Excel in solving the MIVS problem, which is a typical example of hard to solve combinatorial programming problems. Although the Solver, the Excel add-on, cannot be used to optimize problems of this type, end-users could construct optimization models by utilizing other advanced features of Excel. The spreadsheet approach is more attractive to an end-user for a number of reasons. The analytical tools are commonly available and more familiar, the data are entered in a clear and direct way, they are more portable, and their layout is more business like, the results are immediately interpretable, explicit mathematical formulations are not required, and finally the user has access to other office tools such as graphs, and databases to support the problem's analysis.

Another important aspect is the possibility of incorporating new calculations in a simple and transparent way that can be easily explained in order to convince clients and managers about the appropriateness of the approach.

There are technical limitations described in section 4.5 but with computing power of personal computers doubling every few years, restrictions on problem size and computation time will be factors with less significance.

## 6 REFERENCES

1. Akinc, U., Selecting a set of vendors in a manufacturing environment, *Journal of Operations Management*, 11, (1993), 107-122
2. Albers S., Impact of types of functional relationships, decisions, and solutions on the applicability of marketing models, *International Journal of Research in Marketing*, Vol.17, (2000), pp. 169-175
3. Ashley, D. A Spreadsheet Optimization System for Library Staff Scheduling, *Computers and Operations Research*, Vol. 22, No. 6, (1995), pp.615-624.
4. Barlow J.F., Excel models for business and operations management, John Wiley & Sons, West Sussex, (1999), UK
5. Chien Y.I., Cunningham H.J., Incorporating production planning in business planning: a linked spreadsheet approach, *Production Planning & Control*, Vol. 11, (2000), pp. 299-307
6. Chu S.C., Tang B.K.P., An optimization modelling for customer redistribution and its spreadsheet implementation, *Computers and Operations Research*, Vol. 22, (1995), pp. 335-343
7. Conway, D.G., Ragsdale C.T, Modelling Optimization Problems in the Unstructured World of Spreadsheets, *Omega International Journal of Management Science*, Vol. 25, No. 3, (1997), pp. 313-322.
8. Cornford T. and Doukidis G.I., An investigation of the use of computers within OR. *European Journal of Information Systems*, 1, (1991), pp. 131-140
9. Cornuejols, G., Fisher, M.L., Nemhauser, G.L., "Location of bank accounts to optimize float: An analytic study of exact and approximate algorithms", *Management Science*, Vol. 23, No. 8, (1997), 789-810
10. Degraeve, Z., Labro, E., Roodhooft, F., "An evaluation of vendor selection models from a total cost of ownership prospective", *European Journal of Operational Research*, 125 (2000) 34-58
11. Enns S.T., A simple spreadsheet approach to understanding work flow in production facilities, *Total Quality Management*, Vol. 10, (1999), pp 107-119
12. Ferguson, N., Langford, D., Chan, W., "Empirical study of tendering practice of Dutch municipalities for the procurement of civil-engineering contracts", *International Journal of Project Management*, Vol. 13, No. 3, (1995), pp.157-161
13. Fylstra D., Lasdon L., Watson J., Design and use of the Microsoft Excel solver, *Interfaces*, Sept.-Oct., (1998), pp. 29-55
14. Hesse R., Managerial Spreadsheet Modelling and Analysis, Irwin, (1997), USA

15. Hoekman, B., "Using international institutions to improve public procurement", *World Bank Research Observer*, Vol. 13, No. 2, (1998) 249-269
16. Hoekman, B., "Using international institutions to improve public procurement", *World Bank Research Observer*, Vol. 13, No. 2, (1998) 249-269
17. Johnson D.J., A spreadsheet method for calculating work completion time probability distributions of paced or linked assembly lines*, International Journal of Production Research*, Vol. 40, (2002), pp. 1131-1153
18. Novak, D.C., Ragsdale, C.T., A decision support methodology for stochastic multi-criteria linear programming using spreadsheets, *Decision Support Systems*, Vol. 36, (2003), pp. 99-116
19. Osman R.B., Multivariate constrained optimization using PC-based spreadsheet, *Advances in Engineering Software*, Vol. 14, (1992), pp. 137-144
20. Parlar M., Solving dynamic optimization problems on a personal computer using an electronic spreadsheet, *Automatica*, Vol. 25, (1989), pp. 97-101
21. Pearson M.M., Mundell l., Spreadsheet modelling of spatial problems for the classroom, *Decision Sciences Journal of Innovative Education*, Vol. 1, (2003), pp. 133-139
22. Roy A., Lasdon L., Plane D., End-user optimization with spreadsheet models *European Journal of Operational Research*, Vol. 39, (1989), pp. 131-137
23. Thiriez H., Improved OR education through the use of spreadsheet models, *European Journal of Operational Research*, Vol. 135, (2001), pp. 461-476